\documentstyle[prl,multicol,epsfig,aps]{revtex}
\begin{document}

\title{Thermodynamic approach to   
dense granular matter: a numerical realization of a decisive experiment.} 
 
\author{Hern\'an A. Makse$^1$, and Jorge Kurchan$^2$ }  
  
\address{   
$^1$ Levich Institute and Physics Department, City College of New York, 
New York, NY 10031, US \\
$^2$ P.M.M.H., Ecole Sup\'erieure de Physique et Chimie 
Industrielles, 10 rue Vauquelin 75231 Paris, France. 
} 
 
\date{\today} 
 
\maketitle

Years ago Edwards proposed a thermodynamic description of dense
granular matter, in which the grains (the `atoms' of the system)
interact with inelastic forces.  The approach is intriguing but is not
justified from first principles, and hence, in the absence of conclusive
tests of its validity, it has not been widely accepted.  We perform a
numerical experiment with a realistic granular matter model specially
conceived to be reproducible in the laboratory.  The results strongly
support the thermodynamic picture.

\begin{multicols}{2}

The similarities between a driven granular system and a system of
molecules in a fluid have prompted the use of statistical ideas to
describe granular matter \cite{chicago}, 
such as for example kinetic theories of 
gases to study  rapid flows of low-density systems of inelastic
particles\cite{jenkins}. 
In this context, it is customary to associate  the mean kinetic energy of the
particles with a `granular temperature', having no real
thermodynamic meaning.

For the opposite limit of dense, slowly flowing granular matter, a
more radical train of ideas was initiated more than a decade ago by
Edwards and coworkers: the proposal of a statistical ensemble
\cite{Sam,Sam1,mehta,Grinev}, and through it, thermodynamic notions such as
entropy and temperature (the latter unrelated to the `granular
temperature' above).  Although the idea of a thermodynamic description
of granular matter was recognized as attractive, it was not
universally accepted because there is no known first principle
justification of Edwards' statistical ensemble, such as there is for
ordinary statistical mechanics of liquids or gases (Liouville's
theorem).

Recently, analytical developments  and numerical work on
 schematic models originally devised for  glasses have given a  new 
perspective for the dynamics of dense granular matter
\cite{Cukupe,nicodemi,stabarbara,bklm,modelitos1,modelitos2}.
 The  theoretical support provided by this framework 
 for a  statistical approach
has spurred a renewed interest in Edwards' thermodynamics.

On the experimental side, years ago Nowak {\em et al.}~\cite{Sid} studied in
detail the density 
fluctuations in a vibrated granular material, and proposed 
that these fluctuations 
should reflect an underlying  thermodynamics, much as they do in an
ordinary thermal system. 
However, as far as the actual verification of Edwards' approach,  
the evidence they found was if anything rather negative: we shall
discuss below some possible reasons for this.
On the other hand,  Edwards' approach  has never been 
tested with simulations of  realistic models of granular
materials,  characterized by the fact that energy is supplied
by external driving (via shear or tapping) and dissipated by
inelastic collisions and slippage between the grains. 

 In this work we describe a numerical experiment specially
conceived to be reproducible in the laboratory, using a realistic
model of sheared granular matter.  Considering particles of different
sizes inside the medium, we compute a dynamical temperature from an
Einstein relation relating random diffusion and mobility (transverse
to the shear direction) at long time scales.  If there is an underlying
thermodynamics, it will impose that the dynamical temperature be
independent of the tracer's shape: a very strong condition which we
show to hold.
We then perform an explicit computation to show that the temperature
arising from Edwards' thermodynamic description and the dynamic
one measured
from Einstein's relation coincide. This last step cannot be performed
in the laboratory, so the numerical simulation provides the missing
link between thermodynamic ideas and diffusion-mobility checks.



Consider a `tracer' body of arbitrary shape immersed in a liquid in
equilibrium at temperature $T$. As a consequence of the irregular
bombardment by the particles of the surrounding liquid, the tracer
performs a diffusive, fluctuating `Brownian' motion.  The motion is
unbiased, and for large times the average square of the displacement
goes as $\langle |x(t)-x(0)|^2 \rangle = 2 D t $, where $D$ is the 
diffusivity.  On the other hand,
if we pull gently from the tracer with a constant force $f$, the
liquid responds with a viscous, dissipative force.  The averaged
displacement after a large time is $\langle [x(t)-x(0)] \rangle = f
\chi t$, where $\chi$ is the mobility.  
Clearly, the same liquid molecules are responsible both for
the fluctuations and for the dissipation.  Although both $D$ and
$\chi$ strongly depend on the shape and size of the tracer, they turn
out to be always related by the Einstein relation $D/\chi = T$, where
$T$ is the temperature of the liquid.

Conversely, if in a fluid of unknown properties we find that several
tracers having different diffusivities and mobilities yield the same
ratio $D/\chi=T$, we may take this as a strong evidence for
thermalisation at temperature $T$.  Indeed, recent analytic schemes
for out of equilibrium glassy dynamics have suggested the existence of
such a temperature $T$ governing the slow components of fluctuations
and responses of all observables
\cite{Cukupe,bklm}.

In order to test the existence of this temperature for dense granular
matter, we perform, with a realistic numerical model, a
diffusion-mobility experiment in conditions that can be reproduced in
the laboratory.  In the model, deformable spherical grains interact
with one another via non-linear elastic Hertz normal forces, and
non-linear elastic and path-dependent Mindlin transverse forces 
\cite{johnson}.
The normal force, $F_n$, has the
typical 3/2 power law dependence on the overlap between two spheres
in contact, while the transverse force, $F_t$, depends on both the
shear and normal displacements between the spheres.
For two spherical grains with radii
$R_1$ and $R_2$: $F_n = \frac{2}{3}~ k_n R^{1/2}w^{3/2},$ $\Delta F_t
= k_t (R w)^{1/2} \Delta s.$ Here $R=2 R_1 R_2/(R_1+R_2)$, the normal
overlap is $w= (1/2)[(R_1+R_2) - |\vec{x}_1 - \vec{x}_2|]>0$, and
$\vec{x}_1$, $\vec{x}_2$ are the positions of the grain centers.  The
normal force acts only in compression, $F_n = 0$ when $w<0$.  The
variable $s$ is defined such that the relative shear displacement
between the two grain centers is $2s$. The prefactors $k_n=4 G /
(1-\nu)$ and $k_t = 8 G / (2-\nu)$ are defined in terms of the shear
modulus $G$ and the Poisson's ratio $\nu$ of the material from which
the grains are made (typically $G=29$ GPa and $\nu = 0.2$, for
spherical glass beads).
We assume a distribution of grain radii in which $R_1=0.105$ mm for
half the grains and $R_2=0.075$ mm for the other half.  The
observables are measured in reduced units: length in units of $R$,
force in units of $G R^2$, time in units of $\sqrt{\rho R^2/G}$, where
$\rho$ is the density of the particles.  
Internal dissipation is included in two ways: (1) via a viscous
damping term proportional to the relative normal and tangential
velocities, and (2) via sliding friction: when $F_t$ exceeds the
Coulomb threshold, $\mu F_n$, the grains slide and $F_t = \mu F_n$,
where $\mu$ is the friction coefficient between the spheres (typically
$\mu=0.3$).

We perform molecular dynamics (MD) simulations for a binary system of
200 large and small spheres in a periodic 3D cell.  
 Our calculation begins with a numerical
protocol designed to mimic the experimental procedure used to prepare
dense packed granular materials.  The simulations start with a gas of
spherical particles located at random positions in a periodically
repeated cubic cell of side $L$.  The system is then compressed and
extended slowly until a stationary situation with
a specified value of the pressure and volume
fraction--- above the random close packing fraction--- is achieved.
We then apply a gentle shear in the $y$-direction at constant volume
by moving the periodic images at the top and bottom of the cell with
velocities $\dot \gamma L/2$, where $\dot \gamma$ is the shear rate
(Lees-Edwards boundary conditions).
In this simple shear flow the gradient of the
velocity along $z$ is uniform (no shear bands).  
We apply a shear rate of $\dot\gamma = 10^{-4}$, which allows the
system to be in the quasi-static regime.  We focus our study on the
slow shear rate, quasi-static limit where the system is always close
to jamming, but moves just barely enough to avoid stick-slip motion
\cite{savage}. Here the external pressure is large enough
($\sim 10$ MPa in a typical simulation) and the average coordination
number is high enough (typically $\sim 7$) that deformation and
elasticity of the particles play the dominant role, as opposed to the
collision dominated rapid flow regime described by kinetic theories.
We checked that shear induced segregation is absent at the times
scales of our simulations.
The contacts between particles are enduring and the internal 
stresses in the
system are transmitted via a network of `force chains'
\protect\cite{force_chains,chicago}.
independent of the shear rate.

\begin{figure}
\centerline{ 
\epsfxsize=7.cm 
\epsfbox{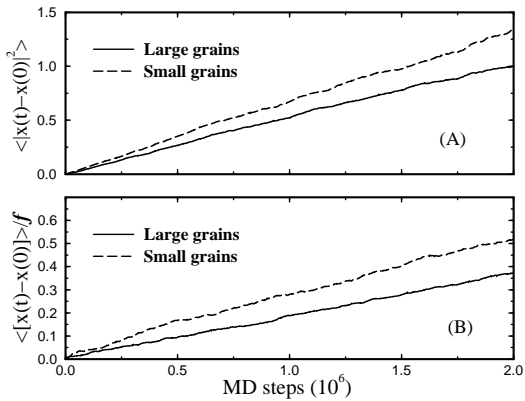}
} 
\narrowtext
\caption{(A) Diffusion and (B) response function for the large and
small particles in a sheared granular material measured perpendicular to the 
shear plane as a function of time in MD steps. 
Both quantities depend linearly on time at the late stages of the 
evolution. 
The response function is measured by applying a constant force $f$ in the 
$x$-direction to each  type of particle. 
Averages are taken over 30 temporal realizations.
The obtained diffusivities and mobilities depend on the 
particles size as expected.}
\label{DX}
\end{figure}

\begin{figure}
\centerline{
\hbox{
\epsfxsize=7cm \epsfbox{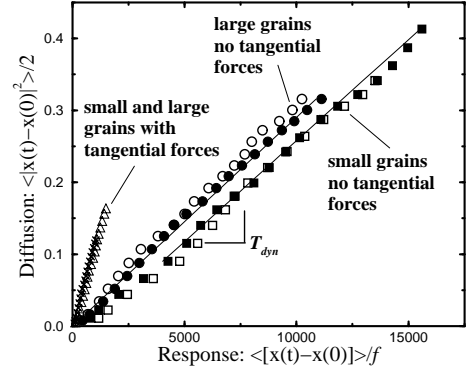}
}
}
\narrowtext
\caption{Parametric plot of diffusion vs  response function  for
small and large grains and for spheres interacting with tangential
forces and without tangential forces (Coulomb frictionless).  
The fitting at long time scales shows the existence
of a well defined temperature which is the same for small and large
grains: $T_{dyn}=2.8 \times 10^{-5}$ for grains without transverse
forces and $T_{dyn}=1.2 \times 10^{-4}$ for grains with Mindlin
transverse forces and Coulomb friction. 
We
calculate the response function for several small external fields and
find the same temperature indicating that we are in the linear
response regime.  Plotted are results  for a
system  without transverse forces
using: $f=1.7 \times 10^{-5}$ (small grains $\Box$,
large grains $\circ$) and $f=2.6 \times 10^{-5}$ (small grains 
$\protect\rule{2mm}{2mm}$, 
large grains $\bullet$). For a system with tangential forces and Coulomb
friction we show the case $f=6 \times 10^{-5}$
(small grains 
$\bigtriangleup$, large
grains $\times$).}
\label{CK}
\end{figure}

After a transient of
the order of the inverse shear rate, we start measuring the
spontaneous $\langle |x(t)-x(0)|^2 \rangle$ and force-induced
displacements $\langle [x(t)-x(0)] \rangle/f $ along the $x$-direction
for the two types of particles with different sizes.  The results are
shown in Fig. \ref{DX}.  We notice that the diffusivities and the
mobilities are different for the two type of particles, since they
have different sizes.  However, when we draw the parametric plot of
$\langle |x(t)-x(0)|^2 \rangle$ versus $\langle [x(t)-x(0)] \rangle/f
$ (Fig. \ref{CK}) we find parallel straight lines for large time
scales, implying an extended Einstein relation:

\begin{equation}
\langle |x(t) -x(t_w)|^2\rangle = 2 ~ T_{dyn} \frac{\langle
[x(t)-x(t_w)]\rangle}{f},
\label{tdyn}
\end{equation}
valid for both particles {\em with the same $T_{dyn}$} for widely
separated time scales $t\gg t_w$. This suggests that $T_{dyn}$ can
be considered to be the temperature of the slow modes.

For small time scales (fast rearrangements, near the origin 
 in Fig. \ref{DX}) the
fluctuation-dissipation plot shows that there is not  a well defined
Einstein-relation temperature. 
We expect this result since the fast motion of the grains 
depends on the microscopic interactions dominated by inelastic
collisions \cite{Foot}.
We also calculate the `granular kinetic temperature'
defined in terms of the velocity fluctuations in the
$x$-direction. Unlike $T_{dyn}$, we find that this temperature is
different for the two types of particles, and their values are two
orders of magnitude smaller than $T_{dyn}$.  We expect this result since the
kinetic granular temperature is dominated by the fast (high frequency)
 modes,  which  are not thermalised to $T_{dyn}$.

We also repeat the numerical experiment for a system of Hertz spheres
without transverse forces ($\mu=0$: experimental realizations of
 elastic spheres with viscous forces
but without sliding friction are foams and compressed emulsions
\cite{bubbles1,bubbles2}) and find that $T_{dyn}$ is well
defined at long time scales for this case as well (see Fig. \ref{DX}).
Thus, our results suggest that the validity of a  structural
temperature for long-scale displacements (larger than a fraction of
the particle size) holds in the presence of viscous forces between
grains or even of a sliding threshold (Coulomb's law).

The existence of a single temperature is an instance of the zero-th
law of thermodynamics, for which we find positive evidence here,  at variance
 with the experimental result in Nowak {\em et al} \cite{Sid}.
 Three possible reasons for the apparent violation
 of the `zero-th law' in their experiments are {\em i)} the effect of an
unknown height-dependent pressure (as pointed out by authors), 
{\em ii)} a rather high tapping amplitude (Edwards ensemble is
in principle only valid in the high compaction limit \cite{bklm}) and
{\em iii)} the fact that the density fluctuations considered are
integrated over all frequencies, thus including also `fast' relaxations.

Next, we treat the question whether it is possible to relate the
dynamical temperature obtained above to the thermodynamic construction
of slowly driven out of equilibrium systems proposed by Edwards.
Whereas in the Gibbs construction for equilibrium statistical mechanics
one assumes that the physical quantities are obtained as average over
all possible configurations, Edwards ensemble consists of only the
blocked configurations (static or jammed) at the
appropriate energy and volume.
The strong  `ergodic' hypothesis is that all blocked configurations 
of given volume and energy 
can be taken to have equal statistical weights.  
This formulation leads to an entropy $S_{Edw}(E,V)$,
and the corresponding temperature $T^{-1}_{Edw}=\partial
S_{Edw}/\partial E$ and compactivity $X^{-1}_{Edw}=\partial
S_{Edw}/\partial V$.

In order to calculate $T_{Edw}$ and compare with the obtained
$T_{dyn}$ we need to count the number of blocked configurations at a
given energy and volume.  (For this calculation we concentrate in the
case without tangential forces and sliding friction, in order to avoid
path dependency which would lead to an ambiguity in the definition of
blocked configurations---see below).  To do this in practice, we
extend the `auxiliary model' method \cite{bklm} to the case of
deformable grains. We define an auxiliary model composed of the true
deformation energy ($E$, the Hertzian energy of deformation of the
particles) and of a term that vanishes in (and only in) the blocked
configurations: $E_{block} \propto \sum_a \left|\vec{F}_a\right|^2$.
(Here $\vec{F}_a$ is the total contact force exerted on particle $a$
by its neighbours).  We perform equilibrium MD with two auxiliary
temperatures $(T^*,T_{aux})$, corresponding to the partition function
$\sum \exp[- E/T^* - E_{block}/T_{aux}]$. Annealing $T_{aux}$ to zero
selects the blocked configurations ($E_{block}=0$), 
while $T^*$ fixes the energy $E$.
We start by equilibrating the system at high temperatures ($T_{aux}$
and $T^*$ $\sim \infty$) and anneal slowly the value $T_{aux}$ to zero
and tune $T^*$ so as to reach the observed value of $E$ during shear.
At the end of the process ($T_{aux}\to 0$),
$T^*(E)=T_{Edw}(E)$, since in this limit we are sampling the
configurations with vanishing fraction of moving particles at a given
$E$.

\begin{figure}
\centerline{
\hbox{
\epsfxsize=7cm \epsfbox{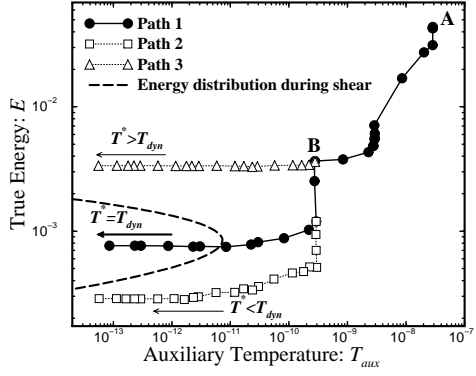}
}
}
\narrowtext
\caption{Annealing procedure to calculate $T_{Edw}$ at different true
energies.  We plot the true deformation energy versus $T_{aux}$
together with the distribution of deformation energies obtained during
shear (dashed curve, mean value $\langle E \rangle = 8.4\times
10^{-4}$).
We equilibrate the system for 40 million iterations at A:
$(T^*=3.4\times 10^{-2}, T_{aux}=3\times 10^{-8})$. We then anneal
slowly both temperatures until B: $(T^*=3.4\times 10^{-4},
T_{aux}=3\times 10^{-10})$, where we split the trajectory in three
paths in the $(T^*,T_{aux})$ plane.  Path 1: we anneal $T_{aux}\to 0$
and $T^* \to 2.8\times 10^{-5}$ which corresponds to $T_{dyn}$
obtained during shear (Fig. \protect\ref{CK}).  Path 2: we anneal
$T_{aux}\to 0$ and $T^* \to 3.4\times 10^{-6}$.  Path 3: we anneal
$T_{aux}\to 0$ but keep $T^*=3.4\times 10^{-4}$ constant.  When we set
$T^* = T_{dyn}$ (Path 1), the final true energy value when $T_{aux}\to
0$ falls inside the distribution of energies obtained during shear and
it is very close to $\langle E \rangle$. This proves that $T_{dyn} =
T_{Edw}$ under the numerical accuracy of the simulations.  For other
values of $T^* \ne T_{dyn}$ the final $E$ falls out of the
distribution obtained during shear (Paths 2 and 3).
We also follow different trajectories (not shown in the figure)
to $T^* \to 2.8\times 10^{-5},T_{aux}\to 0$  and find the same
results indicating that our procedure is independent
of the annealing path.
}
\label{annealing}
\end{figure}

Figure \ref{annealing} shows various annealing protocols. We plot the
value of the true deformation energy as a function of $T_{aux}$ during
annealing using different values of $T^*$ as indicated in the
figure. At the end of the annealing process the system stabilizes at a
given deformation energy.  Only when $T^*$ is equal to $T_{dyn}$
obtained through the Einstein relation during shear, 
we find that the final $E$
coincides, within the accuracy of the simulations, with the mean
energy of the system under shear. For other values of $T^*$ the
annealed system equilibrates at energies which are above of below the
distribution of deformation energies during shear as seen in
Fig. \ref{annealing}.  {\em This proves the equality between Edwards' and
the dynamic temperature.}

We conclude with some remarks: {\em i)} Since the blocked
configurations are the same whatever the inter-grain dissipation
coefficient, Edwards' ensemble (and hence its temperature) are {\em
insensitive to viscous dissipation}.

{\em ii)} On the contrary, tangential forces and sliding friction
block certain configurations, depending on how they are accessed.
The ensemble of blocked configurations is then ill-defined. We have
not tried to construct any ensemble, but content ourselves with the
observation that $T_{dyn}$ is also in this case independent of the
particle size (Fig. \ref{CK})--- we suspect that thermodynamic
concepts apply, but that the relevant ensemble goes beyond Edwards'
construction as it stands.

{\em iii)} We have tested the validity of the thermodynamics in an
ideal homogeneous system with periodic boundary conditions by
explicitly avoiding structural features of dense granular flows such
as shear bands and segregation of the species.  
Even though it remains to be seen whether  the
thermodynamic picture can account for these inhomogeneous
 effects, our ideal system may prove to be useful in
deriving constitutive relations to be used in macroscopic theories of
slow granular flows.

To summarize: we have performed numerically an experiment with dense
granular systems specially conceived to be a `dress rehearsal' for the
real laboratory one. The independence of the Einstein-relation
temperature on the tracer provides a strong test for the thermodynamic
ideas.  We have also showed that this temperature, obtained from
measurable quantities, indeed matches Edwards' one.  If, as now seems
likely, this result is confirmed in the laboratory, this will give an
experimental foundation for the use  of 
the powerful tools of statistical mechanics as
the framework to study this kind of far-from-equilibrium dissipative
dynamical system.

\pagebreak

\end{multicols}
 

\begin{thebibliography}{99}
\bibitem{Foot}  This difference between fast and slow relaxations
 is similar
to the one found for glasses, although in the case of glasses 
the fast modes are equilibrated at the bath temperature \cite{Cukupe}.
\end{thebibliography}
\end{document}